\documentclass[fleqn,10pt]{wlscirep}
\usepackage[utf8]{inputenc}
\usepackage[T1]{fontenc}
\usepackage{filecontents, pgffor}
\usepackage{subcaption}
\usepackage{graphicx}
\usepackage{chemformula}

\usepackage{todonotes}

\newcommand{\lam}[1]{\textcolor{black}{#1}}
\newcommand{\lamrv}[1]{\textcolor{black}{#1}}
\newcommand{\arne}[1]{{\textcolor{black}{#1}}}
\newcommand{\pp}[1]{\textcolor{black}{#1}}
\newcommand{\rv}[1]{\textcolor{black}{#1}}

\title{Analysis of Distributed Ledger Technologies for Industrial Manufacturing}

\author[1*+]{Lam Duc Nguyen}
\author[2+]{Arne Bröring}
\author[3]{Massimo Pizzol}
\author[4]{Petar Popovski}

\affil[1]{CSIRO Data61,  Sydney 2015, Australia. }
\affil[2]{Siemens AG, Munich 81739, Germany.}
\affil[3]{Department of Planning, Aalborg University, Aalborg 9220, Denmark.}
\affil[4]{Connectivity Section, Aalborg University, Aalborg 9220, Denmark. }
\affil[*]{Email: lam.nguyen@data61.csiro.au}
\affil[+]{These authors contributed equally to this work}

   
\begin{abstract}

{In recent years, industrial manufacturing has undergone massive technological changes that embrace digitalization and automation towards the vision of intelligent manufacturing plants. With the aim of maximizing efficiency and profitability in production, an important goal is to enable flexible manufacturing, both, for the customer (desiring more individualized products) and for the manufacturer (to adjust to market demands). Manufacturing-as-a-service can support this through manufacturing plants that are used by different tenants who utilize the machines in the plant, which are offered by different providers. To enable such pay-per-use business models, Distributed Ledger Technology (DLT) is a viable option to establish decentralized trust and traceability.
Thus, in this paper, we study potential DLT technologies for efficient and intelligent integration of DLT-based solutions in manufacturing environments. We propose a general framework to adapt DLT in manufacturing, and then we introduce the use case of \emph{shared manufacturing}, which we utilize to study the communication and computation efficiency of selected DLTs in resource-constrained wireless IoT networks.}

\end{abstract}

\begin{document}

\flushbottom
\maketitle

\section*{Introduction}

% manufacturing is changing:
Industrial Internet of Things (IIoT) is a recent concept that gained traction with the emergence of the wireless 5G technology and it is already exhibiting a great impact within the manufacturing domain\cite{kafle2021topcon}. The general trend is to embrace digitalization and automation towards manufacturing plants that act as cyber-physical systems. This results in an increasing number of smart devices with sensors and actuators that are being integrated in industrial automation processes. In parallel, local edge computing infrastructures are being built up in manufacturing plants, which provide resources for advanced computing and henceforth the basis for next generation IIoT applications\cite{soret2021learning}.
The key economic driver behind this technological evolution is the increase in the production flexibility. This allows for smaller lot sizes and more individualized products for customers. These trends are supported by business models, such as  manufacturing-as-a-service, where manufacturing facilities are utilized more flexibly by numerous tenants who utilize the machines in the plant, which are offered by different providers.
% more complex situation -> trust needed:
These economic forces drive manufacturing plants towards an increase in technological complexity and require improvements in system reliability, intelligence, and trustworthiness during operation\cite{helu2015identifying}. 
Especially the opening of the manufacturing plant's ecosystem to a diverse set of involved parties poses many challenges for manufacturing enterprises to satisfy the trust requirements of multi-partner collaboration \cite{al2019blockchain}.

% blockchain can be used in manufacturing, explain blockchain / DLT:
Distributed Ledger Technology (DLT) can be used to address those trust and privacy challenges in the manufacturing environment of the future, e.g., to transparently store machinery's usage data as a basis for pay-per-use business models on the manufacturing shop floor.
A DLT is a distributed ledger of transactions—rather than being kept in a single, centralized location, the information is held by all the nodes of a network\cite{tschorsch2016bitcoin}. In general, all these network nodes have copies of the same ledger. This removes the need for a third-party to assure that rules are being implemented correctly, instead, this is implicitly done through a decentralized system. Although the most widely known instance of DLT is Blockchain, and, specifically, the most popular decentralized digital currency based on Blockchain is Bitcoin , the transactions on a DLT do not have to be financial. In essence, a transaction simply represents a change in state for whichever data point the DLT’s stakeholders want to track. DLTs are driven by consensus: when a node or a \emph{DLT-client} initiates a transaction, its details are broadcast to the entire network, checked by other nodes and accepted if there is consensus. \emph{DLT-clients} are considered as lightweight devices which have limited resources and just initiate transactions, as well as transmit transactions  to \emph{DLT-managers} to validate. \rv{When the DLT takes the form of a Blockchain}, once a transaction has been validated, it is bundled with other transactions into a block of data. Each block is secured via a cryptographic algorithm. This results in a unique signature for each block known as a hash. These blocks are then ordered sequentially into a chain of blocks, with each block also containing the previous block’s hash\cite{alrebdi2022svbe}. This makes it extremely difficult to tamper with a block, as altering a single piece of data would result in a different hash value, making it evident to the DLT’s users and causing the transaction and block to be rejected. 

%\todo{Arne: include above a description of DLT client and server differentiation.}

% benefits of DLT:
In short, DLTs allow the storage of transactions in immutable records and every record is distributed across many nodes. Thus, security in DLTs comes from the decentralized operation, but also from the use of strong public-key cryptography and cryptographic hashes\cite{tschorsch2016bitcoin}. 
The key benefits of the integration of DLTs into manufacturing systems are: 
i) auditable and guaranteed immutability and transparency for stored data (e.g., machine usage data, sensing data about machine conditions, or logs about user/technician engagements), 
ii) no need for a third party to assure the rules between the different parties in the manufacturing ecosystem are met,
%iii) development of a transparent system for heterogeneous resource-constrained wireless IoT data networks to prevent tampering and injection of fake data from the involved untrusted parties\cite{}.
iii) enabling security and privacy of information in manufacturing networks in conjunction with others techniques e.g, private transactions and private channels, which is urgently needed as more than 25\% of cyber attacks will involve IoT\cite{Top10Sec35:online}.

% application:
To showcase these benefits and to have a realistic use case as an example for our studies, we implement in this work the DLT-based application of \emph{shared manufacturing}, which relates to the economic driver of 'flexible production'. Specifically, a robot arm, as part of a production cell with multiple machines, is offered by a provider, who allows different tenants of the plant the usage of the robot arm, while expecting a usage fee. In this application, the DLT is required to capture the usage times of the robot arm through the various tenants, which is then the basis for a correct billing and payment for usage time.
Some parts of this process can be done automatically with smart contracts. These involve two entities turning a business contract into code that recognizes actions on the DLT. For example, a smart contract might recognize that a rental of a machine from “provider A” to “customer B” on a certain date for a specific time period should be for a specific price\cite{nguyen2021marketplace}. This simplifies processes that take significant time to check. This structure gives DLT participants confidence in their transaction without the need to trust each other. Nor do they need to agree on a trusted third party to make sure they’re both following the rules. Because the ledger of transactions is consensus-based and distributed, records stored in it cannot be erased or changed.

% limitations:
To be able to implement the above described application and reap the described benefits, the system designs of current manufacturing plants need to be adjusted to be able to accommodate the operation of a DLT and overcome certain limitations:
% limitations in computation infrastructure:
First, today's computation infrastructures of industrial manufacturing plants are typically designed as centralized systems, where cloud services perform data aggregation and analysis\cite{chen2018edge}. While the manufacturing infrastructure comprises a multitude of IoT devices and sensors that collect data and have only little computing power, the gathering and processing of data in a centralized cloud service may lead to network overload and single points of failure \cite{hassan2019current}. To setup a DLT network in such an environment, a sufficient amount of local computing capacity\cite{shi2016edge} needs to be available and, potentially, edge computing facilities can be integrated in the computation infrastructure.
% limitations in communication infrastructure:
Furthermore, industrial communication systems have been traditionally designed for reliable operation in a noisy factory environment, employing mainly wired and proprietary communication technologies to connect sensors, actuators, and controllers. Nevertheless, with the emergence of IIoT, future factories will increasingly rely on diverse communication technologies, including wireless standards, to ensure reliability, interoperability, and remote operation and control of production processes through the Internet. These wireless links are potentially less reliable and are more constrained, which needs to be considered, when operating a DLT network.
% research question:
\rv{From these limitations regarding the system infrastructure, we derive the key \emph{research question} of this work: \emph{"What is the  computation and communication overhead that results from the operation of a DLT network in a manufacturing environment?"} The answer to this question will be critical to understand for future research on applications of DLT in manufacturing, as well as for practitioners who want to deploy a DLT network in a manufacturing plant.}

%%% RELATED WORK:

The use of DLT in manufacturing has received attention from both academia and industry because of its promise for easing supply chain and manufacturing operation management problems due to its advantages in transparency, traceability, and security \cite{}.
In industry, Bosch increasingly connects their products to the IIoT in order to directly participate in the digital economy. The goal is to build an \emph{Economy of Things}, which will be based on DLT \cite{Economyo60:online}. Another example is a concrete solution by Siemens, which enables their \emph{Mindsphere} IIoT platform to track products of the food and beverage industry transparently throughout their entire life cycle based on DLT \cite{Blockcha24:online}. \emph{MindSphere} exploits all useful information before forwarding only a crucial subset to the distributed ledger. The DLT then makes sure the collected data is safe and transparently accessible to everyone who is part of the ecosystem.
In academia, Li et al. \cite{li2018toward} introduced a distributed P2P system that improves the security and scalability of the cloud-based manufacturing platform based on DLT. Danzi et al.\cite{8671694} analyze the communication aspects in terms of delay and overhead between IoT devices and Blockchain network. The authors demonstrate that, if the statistics of account updates and the channel state are known, the lightweight IoT clients can construct a list of events of interest that provides a predictable average communication cost. In addition, a survey\cite{9129732} about performance of different Blockchains is conducted, but the work mainly focuses on theoretical aspects, and lacks a detailed analysis in specific application areas such as manufacturing.
Fu et al. \cite{fu2018blockchain} presented an innovative environmentally sustainable DLT-energized strategy for the fashion apparel manufacturing industry. 
Yu et al. \cite{yu2020blockchain} proposed a DLT-based service composition architecture for manufacturing. In general, the public  DLT-based applications are characterized by the distinctive metric of computational trust. 

%\todo{Arne: related work above is a bit thin.}
%\todo{Arne: can we include your colleagues work in related work? \url{https://arxiv.org/pdf/1807.07422.pdf}} 

In order to be able to answer our research question, we extend state-of-the-art through the following research contributions: 
\begin{itemize}
    \item General analysis of different DLTs and their capabilities when used in industrial manufacturing environments; 
    \item System design for DLT-based IIoT manufacturing systems that can integrate and adapt multiple features and components;
    \item Evaluation of communication and computation overhead of different DLTs in resource-constrained IoT networks. This benchmark of different DLTs for manufacturing scenarios will help interested parties to understand the trade-offs in DLT-based systems.
\end{itemize}

%\todo{Arne: structure description needs update in the end.}
The remainder of this paper is organized as follows. In the next section, we present the results of this study. First, we present a general analysis of five different DLT platforms. Second, we introduce a system design for using DLT in industrial manufacturing. Third, we implement  the shared manufacturing use case and perform a performance evaluation of the five different DLTs.
Finally, we discuss our findings and indicate avenues for future research.

\section*{Results}
In this section, we first study five different DLT platforms, then we propose a general framework to integrate DLT in manufacturing. Finally, we implement the use case of shared manufacturing and conduct the evaluation.

\subsection*{Analysis of DLTs for Industrial Manufacturing}

Although a large number of DLTs are available, within the scope of whic work we have selected five representative DLT platforms that are either already used or appear as most promising for manufacturing environments: Hyperledger Fabric, Ethereum, Quorum, Solana, and IOTA. 
%\todo{PP: Would be good to have a reference which of these is already used in some of the manufacturing cases described in the previous section.}
%The evaluation and arguments in this research are based on the experimental results in a private DLT network setup at the Siemens research labs. 
%\todo{The specs shown in Table 1 are from literature, right? Any results based on experimental works are shown in the section 'Analysis of Shared Manufacturing Use Case', right? or do you include those results here in the table as well? - The table just shows the general comparison from literature, and next section, we show the experimental results.}
The overview comparison of these DLTs is shown in Table \ref{tab:comparison}. 
%\todo{Please add latency numbers to the table}

%\input{table-comparison}

%%%%%%%%%%%%%%% Table of Comparison

\begin{table}[t!]
\centering
\caption{Comparison of different enterprise DLT platforms}
\label{tab:comparison}
\begin{tabular}{ p{2.3cm} ||p{2.5cm} p{2.5cm} p{2.7cm}  p{2.5cm} p{2.5cm}}
\toprule
& \textbf{Hyper. Fabric} \cite{androulaki2018hyperledger}   & \textbf{Quorum} \cite{baliga2018performance}      & \textbf{Ethereum} \cite{vujivcic2018blockchain}  & \textbf{IOTA} \cite{popov2019iota} & \textbf{Solana}\cite{yakovenko2018solana} \\ 

\midrule
%                                Fabric              Quorum                 Ethereum            IOTA                  Solana
\textit{DLT type}                & Private           & Private              & Public / Private  &  Public / Private   &  Public / Private   \\ 
\textit{Goals}                   & Open DLT framework & Open, based on Ethereum & Broad ecosystem   & Lightweight &  High scalability      \\
\textit{Application}             & Enterprise DLT    & Enterprise DLT       & DApps             &  IoT                &  DApps     \\ 
\textit{Governance}              & Linux Foundation  & ConsenSys            & ETH Foundation    &  IOTA Foundation    &  Solana   \\
\textit{Cryptocurrency}          & N/A                & N/A                  & Ether (ETH)       &  MIOTA              &  SOL            \\
\textit{Consensus}               & Pluggable         & Voting Protocol      & PoW               &  Tangle             &  PoH          \\
\textit{Smart Contract}          & nodejs, go, java  & Solidity             & Solidity    &   {Solidity, Go, Rust}        &  Rust    \\
\textit{Throughput}              & $\sim$2000 tps    & $\sim$100 tps        & $\sim$100 tps     &  1000$\sim$1500 tps &  $\sim$1400 tps     \\
\textit{Latency}   & $\sim$250 ms      & $\sim$414 ms           & $\sim$2150 ms        & $\sim$ 258 ms           & $\sim$ 500 ms      \\
\bottomrule
\end{tabular}
\end{table}

%%%%%%%%%%%%%%%%%%%%

%The table compares the 5 DLT platforms. regarding type, industry focus, governance, availability of an associated cryptocurrency, consensus mechanism, and supported application types.

% public, private, hybrid DLTs:
Each DLT can be categorized as public, private, or hybrid, where the latter one can support features of both public and private ones.
%\todo{Is any of the ones listed here hybrid?}. Public (or permissionless) 
\emph{DLTs} allow any user to pseudo-anonymously join the DLT network and do not restrict the rights of the nodes on the network. 
We are investigating in this paper the public DLTs Ethereum\cite{vujivcic2018blockchain}, IOTA\cite{popov2019iota}, and Solana\cite{yakovenko2018solana}.
%other examples are Bitcoin\cite{tschorsch2016bitcoin}, Bitcoin Cash\cite{javarone2018bitcoin}, or Monero\cite{miller2017empirical}. 
However, for the implementation of our use case within a manufacturing plant such public DLTs are used in a private deployment by installing local networks.
%\todo{'DLT type' in table (and in this sentence) of Ethereum, IOTA and Solana needs to be changed to 'public / private', right? because we will use them in a private fashion, right?}
In contrast, private DLTs restrict access to their network to certain nodes and may also restrict the rights of nodes on the network. In this paper, we are investigating the private DLT platforms  Hyperledger Fabric\cite{androulaki2018hyperledger} and Quorum\cite{baliga2018performance}. The identities of the users of a private DLT are known to the other users of that private DLT. 
In a Hybrid DLT, every transaction can happen quickly in its own private chain and commits to the public chain only happen as and when necessary, e.g., when public verification is required. This provides the immutable trust from the Blockchain as well as the scaling from private DLTs. Layer 2 solutions and side-chains\cite{sguanci2021layer} are variations of this concept.

Besides their type, the five DLTs have different goals and applications in focus. Hyperledger Fabric and Quorum are both aiming to offer a open foundation for new components to build a broad ecosystem that supports enterprises with various functionalities to deploy their own private DLT. Ethereum has a large community of developers and already an established ecosystem that focuses on decentralized applications (DApps), e.g., for decentralized finance. Solana follows a similar application focus, while aiming for higher scalability than Ethereum. IOTA's focus is on IoT applications and therefore aims to support DLT participants with a small footprint.

% SECURITY DISCUSSION:

% moved the following paragraph up here:

\rv{IIoT applications in the manufacturing environment will involve many stakeholders with different roles, functionalities, and information with access rules, identities and security factors. An important factor to provide security is the support to validate transactions generated by participating nodes. While Hyperledger Fabric and Quorum are suited solely for private setups, Ethereum, IOTA and Solana are designed for public networks, but can also be configured for private purposes. In terms of security and confidentiality, public networks can show certain advantages over private ones, especially if they are able to provide transparency and distributed storage. Besides, the more users a public DLT has, the more secure it is. However, for enterprise use (i.e., also for typical manufacturing scenarios) public DLTs are not ideal as companies deal with highly sensitive data and cannot allow arbitrary users to join their network. Also, private DLTs provide very low or no fees for validation and a faster consensus process. However, a private DLT can be altered by its owners, making it more vulnerable to hacking\cite{wust2018you}. 
Besides, only Hyperledger Fabric supports data confidentially by default; this is done via in-band encryption and guarantees the privacy of data by creating private channels (e.g., to setup for departments within an organization). Therefore, Hyperledger Fabric allows for authorization with trusted Certificate Authority per channel. These features are vital in a trusted IoT system for enterprises.}

Each DLT platform deploys a different consensus mechanism. Ethereum uses the  Proof-of-Work (PoW) consensus that requires involved parties of a network to expend effort solving a mathematical puzzle to prevent anybody from gaming the system. PoW consensus consumes significant computing and energy resources, which is  not suitable for resource-limited systems. Quorum, as an enterprise version of Ethereum, uses a voting-based consensus protocol. This consensus protocol achieves consensus on transactions and key network decisions by counting the number of votes cast by nodes on the networks and not consuming more energy for verification as compared to PoW. IOTA uses ``little'' PoW for preventing spamming attacks. In Ethereum, doing PoW is to receive the power to define the truth, the node with more power can solve the PoW faster and consume more energy. Meanwhile, IOTA use PoW with lower difficulty to prevent spamming and to allow transactions to be attached in the Tangle. \rv{The "little" PoW is still similar to original PoW but with lower difficulty.}
%\todo{Could you extend the IOTA PoW description by a few words to explain what is used in case PoW is not used. So, what does 'little' mean?}.
Hyeperledger Fabric modularized the consensus part among distributed peers in an ordering service\cite{androulaki2018hyperledger}, so that this platform allows users to choose their preferred algorithm, e.g., CFT (crash fault-tolerant) or BFT (byzantine fault-tolerant) ordering. Finally, Solana introduces a new consensus algorithm called Proof-of-History which allows timestamp field to be built into the blockchain itself instead of using values of timestamps as PoW DLTs. \rv{Whereas other DLTs require validators to talk to another in order to agree that time has passed, each Solana validator maintains its own clock by encoding the passage of time in a simple SHA-256, sequential-hashing Verifiable Delay Function (VDF)\cite{boneh2018verifiable}.}

Table \ref{tab:comparison} specifies the smart contract programming language supported by the DLT. Smart contracts act as autonomous entities on the ledger that deterministically execute logic expressed as functions of the data that are written on the ledger. Therefore, smart contracts can be established to have automatic reactions from the DLT network to specific events. For example, in the use case of shared manufacturing, smart contracts can be used for restricting, tracking and payment for the usage of the rented machinery. 
The smart contract feature is currently supported by Ethereum, Solana and Hyperledger Fabric (called 'Chaincode'). \rv{In IOTA, a smart contract is called Quobic which is now deprecated and the new beta version of smart contract is developing\cite{IOTASmar41:online}}. 

Furthermore, Table~\ref{tab:comparison}  states the performance characteristics of the DLTs. These values have been acquired both through our own experiments and the data available in the literature.  
%\todo{this is unspecific. For each number we need to specify how we got it.}
These measures are of vital importance for IoT applications, particularly in manufacturing, where a large number of sensors may generate millions of data points per day. This requires high efficiency of the consensus mechanism, including the way in which transactions are processed by the peers, known as \emph{endorsing peers} in Hyperledger Fabric, \lam{validators in Solana,} and full nodes (peers) in Ethereum. \lam{Specifically, we have Solana, with 600 nodes and around 1000 validators. Currently Solana is hosting around 340 apps\cite{Homepage72:online}. Meanwhile, Ethereum has over 3000 Dapps running on its network.} Regarding latency, the transaction confirmation time must be sufficiently short to avoid queuing in the DLT and to ensure consistency in the ledgers. The confirmation time of an Ethereum transaction in a public network is around $25$ seconds in public networks. This value indicates that consensus over public networks may not be suitable for real-time IoT applications. However, other DLT platforms can achieve  much lower confirmation times \cite{xiao2020survey}. Note that in Table \ref{tab:comparison} the transaction confirmation time is included in the end-to-end latency, which however does not account for the communication latency at the radio access networks.   

Another important performance feature of the DLTs is related to the CPU usage and resulting energy consumption. The idea that DLT technologies and crypto-assets consume an excessive amount of electricity has been at the heart of recent discussions around this technology. The energy consumption of a DLT protocol should not be equated with its environmental footprint. Indeed, many use cases related to DLT technologies and crypto-assets may even contribute to improving the environmental footprint, in particular by using the surplus of decarbonised energy in certain geographical areas where the need for electricity is lower than the level of production\cite{ahl2020exploring}. 
%\todo{this needs refernce(s) to example projects}
In the scope of this work, we study the carbon footprint of the different selected DLT platforms within the local testbed of the shared manufacturing use case. 

%The Siemens hardware SIMATIC IPC427E - Embedded IPC\footnote{https://mall.industry.siemens.com/mall/en/WW/Catalog/Products/10290372?activeTab=productinformationandregionUrl=WW} is used as an edge server running DLT nodes. \todo{move this information to the section below.}

%\todo{The discussion on gas fees below is not applicable here, as we consider only locally installed DLTs, right? maybe we can change here the text accordingly to point out gas fees as a reason why we do not consider public networks?}
The charge of fees to process the transactions, commonly known as \emph{gas} is yet another factor to take into account to select the appropriate DLT. These may greatly increase the operational costs of the network, which negatively impacts the throughput of the DLT. On the one hand, transaction fees pose a problem in massive IIoT scenarios if the generation of a large number of transactions is essential. On the other hand, these fees may contribute to minimizing the amount of redundant transactions generated by the sensors, which in turn offloads the DLT. In industrial manufacturing domain, the required fee for generating transactions within a company or among some cooperative organizational setup may not be suitable. In addition, businesses have always required a reasonable degree of privacy as well as control over their networks, so that publishing the data on a Blockchain is not reasonable and potentially unsafe. Therefore, we consider only private Blockchains for the enterprise scenario.
%Among the considered \emph{DLTs}, Ethereum requires fee and gas for each transaction whereas Hyperledger Fabric and IOTA provide free solutions to exchange transactions

\subsection*{System Design for using DLT in Industrial Manufacturing}

\begin{figure}[ht!]
    \centering
    \includegraphics[width=0.7\linewidth]{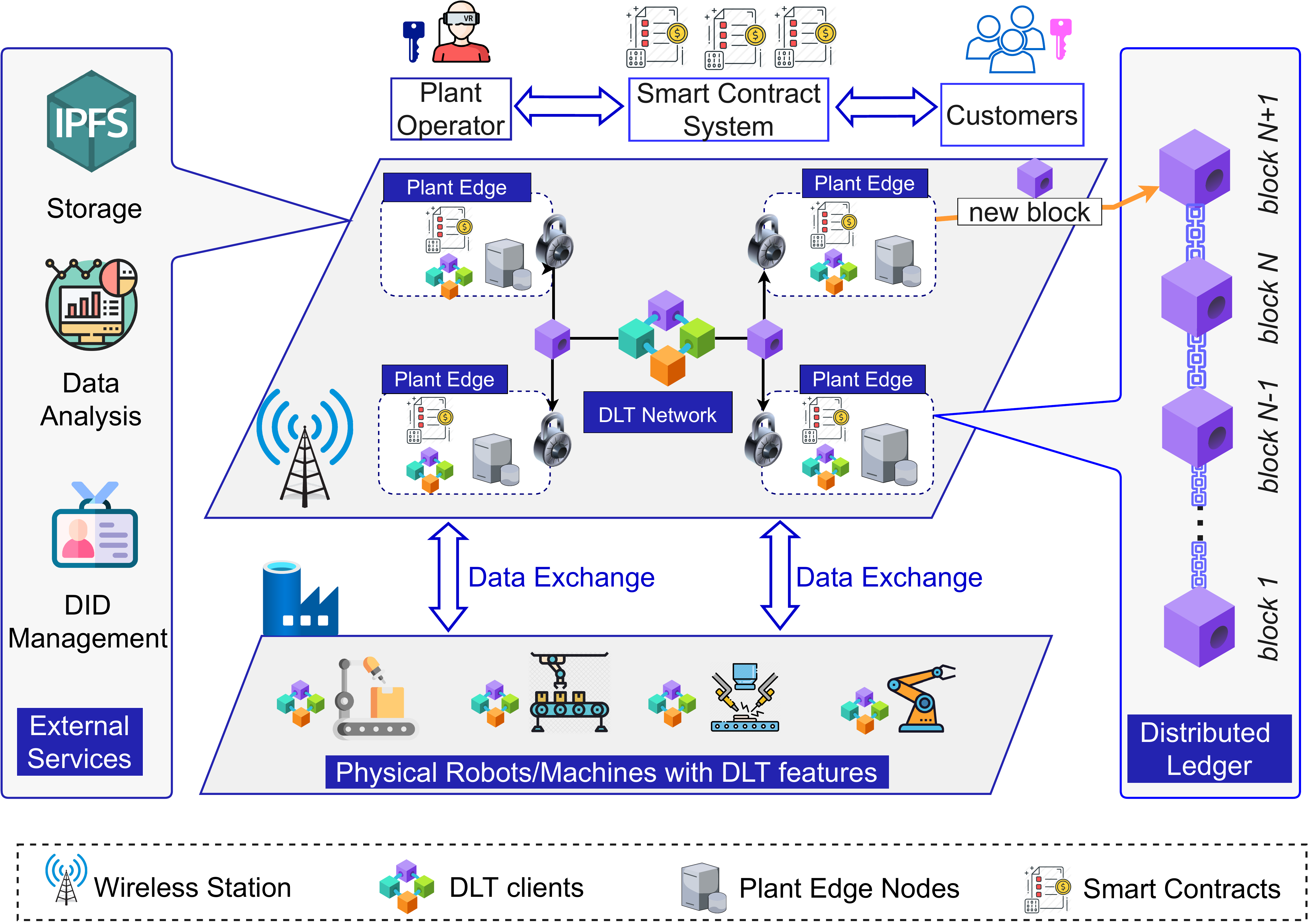}
    \caption{Overview of the system design}
    \label{fig:architecture}
\end{figure}

The proposed system design is described in Figure \ref{fig:architecture}. It comprises four key parts as described below.

%\todo{Arne: Regarding Table 1: the throughput cell for Ethereum states "public". We should present a value for a private Ethereum setting.}

\textbf{DLT System:} This component includes all modules to build various features of DLT technologies such as consensus, smart contract, data authorization, identity management, and peer-to-peer (P2P) communication. These components must ensure that every change to the ledger is reflected in all copies in seconds or minutes and provide mechanisms for the secure storage of the data generated by IoT devices and parameter configurations. There are numerous DLTs with different characteristics that may be beneficial for different target applications. The  DLT nodes can be located everywhere and connected with base stations via the Internet.

\textbf{Physical Machines:} This component consists of physical robots, machines, and IoT sensor devices which collect the data and publish to the distributed ledger for accounting or analyzing purposes. 

\textbf{Plant Edge System:} Even though DLT-based solutions offer significant countermeasures to secure data from tampering and support the distributed nature of the IoT, the massive amount of generated data from sensors and the high energy consumption required to verify transactions make these procedures unsuitable to execute directly on resource-limited IoT devices. Instead, edge servers with high computation resources can be used to handle real-time applications and to further increase the degree of privacy (e.g., through cloud computing)~\cite{xiong2018mobile}. The edge network is a potential entity to cooperate with the DLT network in computationally heavy tasks and return the estimation results \lam{(e.g., from solving proof-of-work (PoW) puzzles, hashing or algorithm encryption)} to the DLT network for verification.

\textbf{External Services:} The devices of the manufacturing environment are typically resource-constrained with limited storage space and low computation capacity. Hence, external infrastructure which operates on the edge may be incorporated to provide external services, such as storage and computing. For example, the Interplanetary File System (IPFS) is a distributed file storage system that can store data generated from IoT networks and return a hash to the ledger based on the content of the data. Since the ledger cannot handle and store the massive amount of manufacturing data collected by the sensors, machines, and robots, the service provided by the IPFS is a vital component. \lamrv{The default configuration of IPFS connects to the global distributed network. In some cases regarding to privacy and confidentiality, a private IPFS network is preferred over connecting to the public IPFS network. In our scenario, we prefer to configure IPFS privately in a local cluster. Second, we introduce the application of payment channels to sharing manufacturing use case because of its natural advantages. In specific, a Payment Channel is a process where customers can make multiple transfers with e.g, plant operator, without sending a transaction to the DLT. Once the final transaction occurs between the participants the recipient can claim their funds by submitting one final transaction to the Smart Contract on the ledger. This allows both parties to avoid fees involved with multiple transactions. Smart Contracts can be an agreement about the rental time, specific tasks between customers and plant operators, or smart contracts created at the beginning of the process of payments.}
In addition, a Digital Identity Management (DID) could be added to support managing identity of participant devices in a distributed manner.
%\todo{Explain also the other external services, e.g., 'DID Management' = Digital Identity Management.}

%The workflow of the proposed system is described as below: 
%\begin{enumerate}
%    \item \textit{Association}. The group of end-devices such as robot arm, gripper associate with plat edge devices via machine-to-machine communication for a specific learning task\cite{}. These end-device nodes are selected by edge nodes that is involved in model training. This big data associated with other smart industries of some other company can also be considered while training the global model\cite{}.
    
   % \item \textit{Distribution}. This initial model is then distributed to end-devices for the purpose of updating and training this model based on clients local raw data. Each training node computes a local model by using its own data and then transmits the local model to its associated plant edge nodes via DLT transactions\cite{}.  
%\end{enumerate}

\subsection*{Performance Evaluation of DLTs in a Shared Manufacturing Use Case}

\begin{figure}[t!]
     \centering
     \begin{subfigure}[b]{0.4\textwidth}
         \centering
         \includegraphics[width=0.9\linewidth]{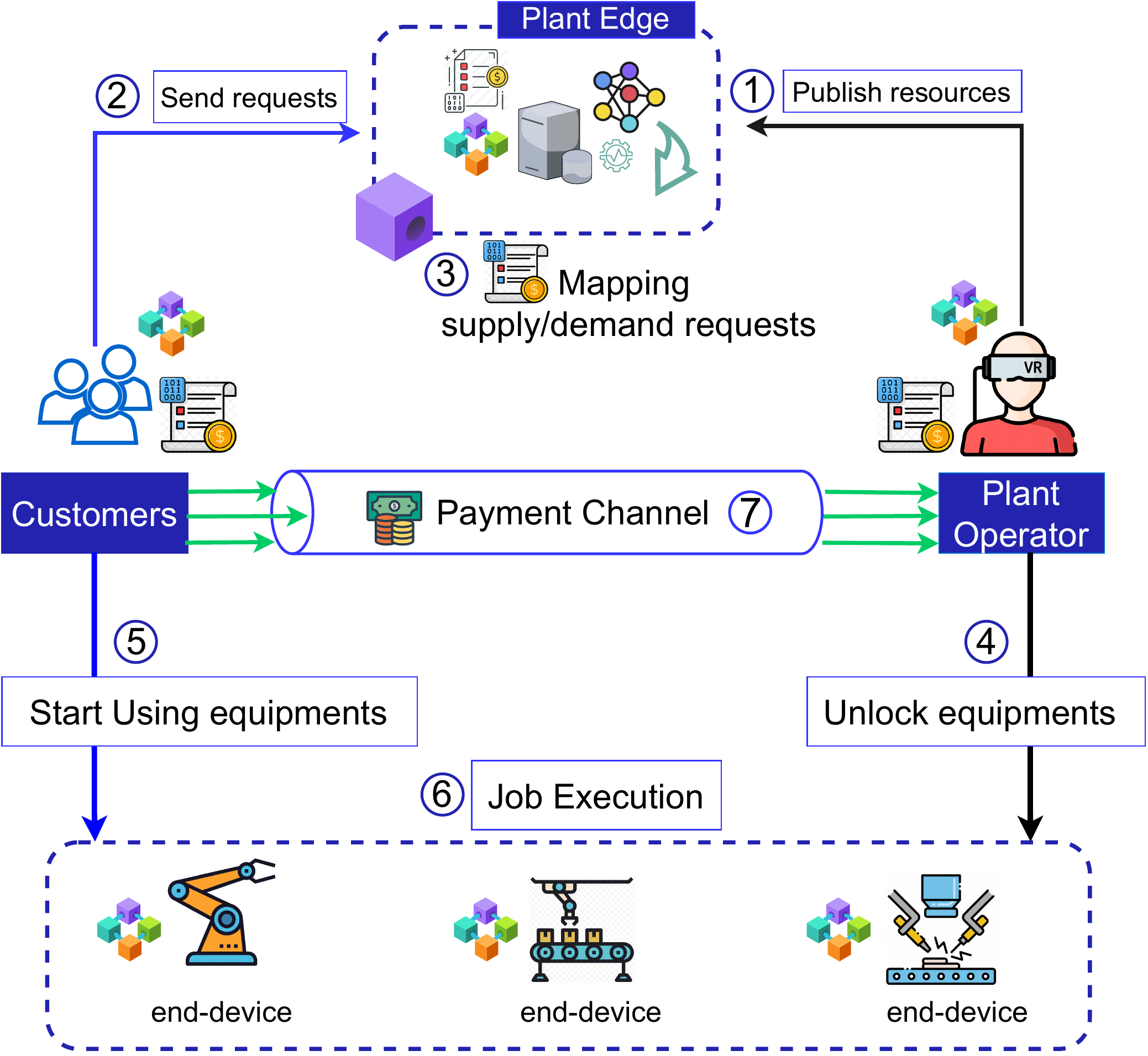}
         \caption{Shared Manufacturing communication workflow}
         \label{fig:example}
     \end{subfigure}
     \hfill
     \begin{subfigure}[b]{0.55\textwidth}
         \centering
         \includegraphics[width=1.0\linewidth]{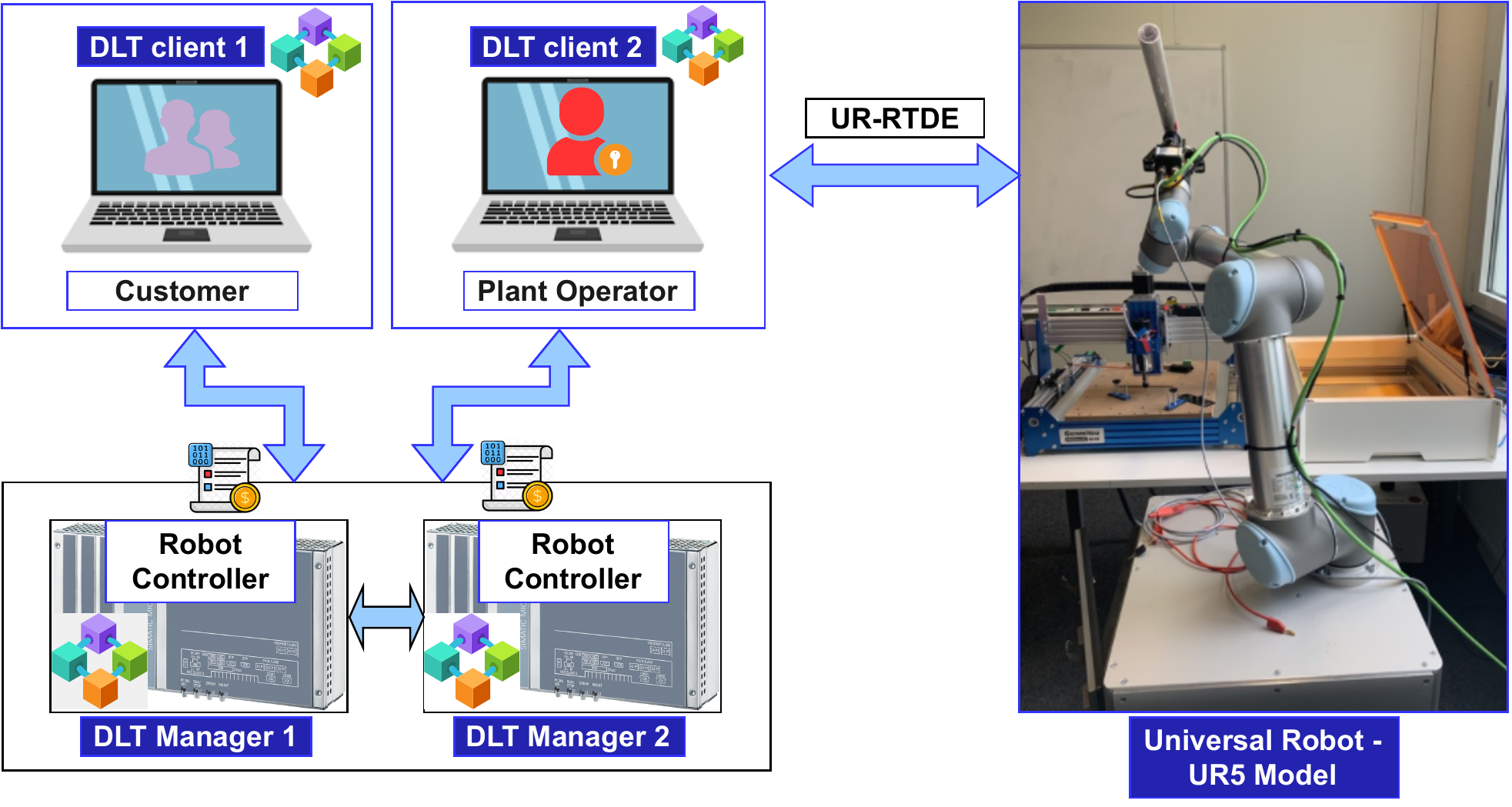}
         \caption{Testbed set up with UR5 robot arm\cite{UR5colla78:online}} %footnote{ \url{https://www.universal-robots.com/products/ur5-robot/}}}
         \label{fig:testbed}
     \end{subfigure}
     \caption{System model of shared manufacturing and local test-bed setup}
\end{figure}

\begin{table}[b]
\centering
\caption{Testbed Settings}
\label{tab:testbedsettings}
\begin{tabular}{ p{2cm}|| p{3cm}| p{3cm}| p{3cm}| p{3cm}  }
\toprule
\textbf{}                   & \textbf{\emph{DLT-manger-1}} & \textbf{\emph{DLT-manager-2}} & \textbf{DLT-Client 1}    & \textbf{DLT-client 2}  \\ 
\midrule
\textit{Devices}    & Siemens Microbox       & Laptop                 & Rapsberry Pi 3+          & Raspberry Pi 3+          \\ 
\textit{RAM}              &     4GB       & 8 GB     & 1 GB & 1GB      \\
\textit{Connectivity} & Ethernet & Ethernet & Wifi & Ethernet \\ 
\textit{Capacity} &           Intel(R) Core i7-351UE CPU @ 1.70GHz x4 GHz             &  Intel(R) Core(TM) i7-8550U CPU @ 1.80GHz   1.9 GHz     & Quad Core 64 bit ARM cortex at 1.2 GHz & Quad Core 64 bit ARM cortex at 1.2 GHz \\ 
\bottomrule
\end{tabular}
\end{table}

In this section, we analyze the application of the \pp{Shared Manufacturing} use case and study its performance. The application uses DLT to automate the management of rentals of industrial robots, where the manufacturing plant operators and their customers can make agreements without third parties and the associated delay.

Along with data sharing\cite{nguyen2021modeling} and vehicle sharing\cite{storch2021incentive}, the machine sharing concept in industry manufacturing has been recently identified as a key innovation for implementing the next industrial evolutionary step\cite{yu2020blockchain}. Open and shared manufacturing factories are composed of a number of industrial robots and other production machines that can be rented by customers. The advantage over traditional manufacturing plants is that such plants can have a higher workload and less idle periods, which in turn can make the production cheaper. Therefore, production tasks need to be efficiently allocated on the available machine resources under consideration of system performance.

Our shared manufacturing application scenario is described in Figure ~\ref{fig:example}. As an initial step (1), the plant operator of a factory publishes the list of machine resources which are available for rent. Thereby, each machine has a unique ID and described capabilities to perform specific jobs. A \emph{manufacturing marketplace} running on a DLT-based network can be implemented in such an environment to offer access to those machine descriptions. In the DLT-based manufacturing network, smart contracts are running to receive requests from customers rent machines (step 2) and match them to resources offered by the plant operator (step 3). In addition, the rules and agreements, e.g., about the rent period, specific tasks, or payment methods between plant operator and customers are pre-defined in the smart contracts and executed autonomously. The customers can check the list of available machines published by the plant operator, and if the customers have a relevant job coming up, they can request the suitable machines via smart contracts. This is the first difference between the DLT-based and non-DLT shared manufacturing system. In a non-DLT based system\cite{RentingM65:online}, a plant operator and customers could not work directly by exchanging messages without the guarantee about the trust of contracts as well as payment. This guarantee requires a third-party to complete the deal. After DLT-based smart contracts executed and mapped the requests from plant operator and customers, the plant operator account will unlock automatically the available machines (step 4) and assign the control of the machines to customers. Then, the customers can start control and program the machines for their jobs (step 5), which are then executing these jobs (step 6). Compared to standard shared manufacturing, the second innovation in DLT-based systems is that we implemented a layer 2 \textit{payment channel}\cite{jourenko2019sok} between the plant operator and customer for micro-payments (step 7).

To study the communication and computation overhead resulting from DLT in manufacturing systems, we have implemented the above described \emph{shared manufacturing} application in a private setup as shown in Figure  \ref{fig:testbed}. The setup involves the DLT components \emph{DLT-manager 1 and 2} and \emph{DLT-clients}. The \emph{DLT-manager} has a high computation capacity and enough storage for a full ledger with all the information and data. The \emph{DLT-clients} are lightweight and are limited in terms of computation and resources. The \emph{DLT-clients} can query and access the data from the ledger without downloading the full chain of blocks. 
%\todo{Please resolve inconsistency regarding the DLT Client: sometimes it is described that we have 1 sometimes 2.}
The \emph{DLT-managers} are implemented in two different equipments: as a Siemens Microbox\cite{UR5colla78:online} and a Macbook Pro. The \emph{DLT-clients} are implemented in Raspberry Pi 3+. The specifications of these devices are found in Table \ref{tab:testbedsettings}. \emph{DLT-manager-1} and \emph{DLT-manager-2} are connected via Ethernet, and communicate with DLT-clients via local WiFi. The communication method can be extended to other long-range communication or global internet depending on specific scenarios.
The distributed ledger is deployed in the DLT Managers. We have implemented five types of \emph{DLTs}, namely Ethereum, Quorum, IOTA, Hyperledger Fabric,  and Solana.
%\todo{Note which plugin consensus is used for Fabric}

\begin{figure}[t!]
     \centering
     \begin{subfigure}[b]{0.85\textwidth}
         \centering
         \includegraphics[width=1.0\linewidth]{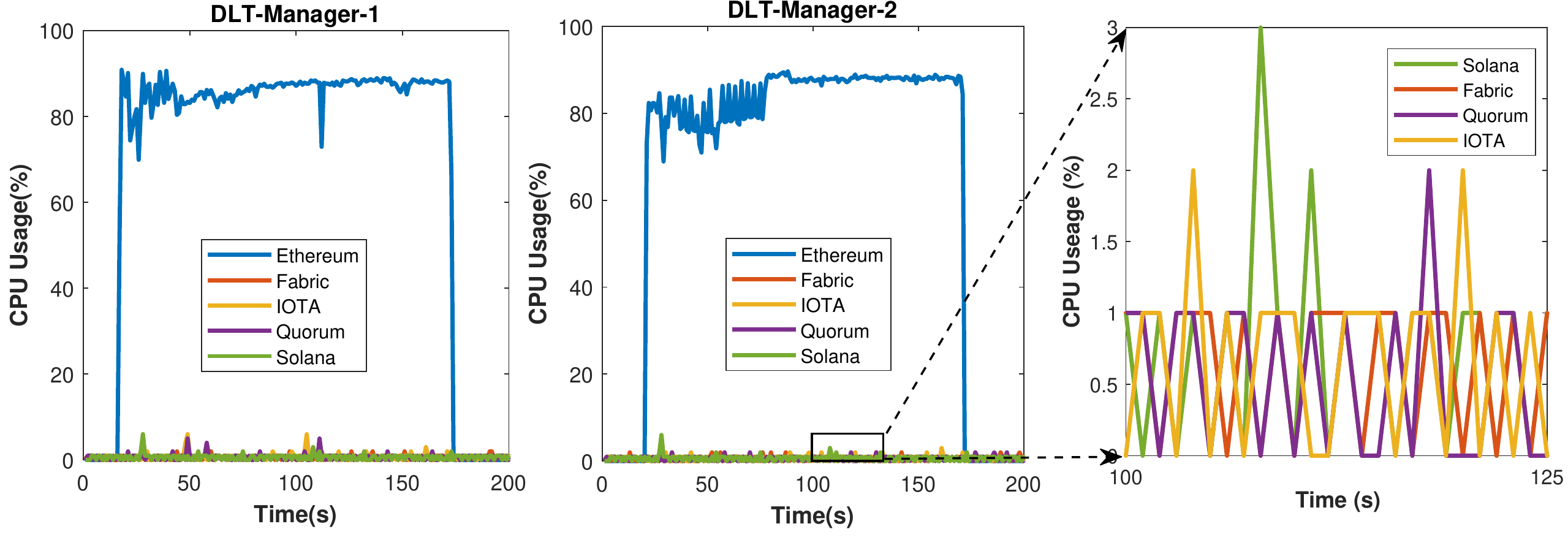}
         \caption{Energy Consumption of \emph{DLT-managers} running full DLT features in separated scenarios.}
         \label{fig:energy-managers}
     \end{subfigure}

     \begin{subfigure}[b]{0.85\textwidth}
         \centering
         \includegraphics[width=1.0\linewidth]{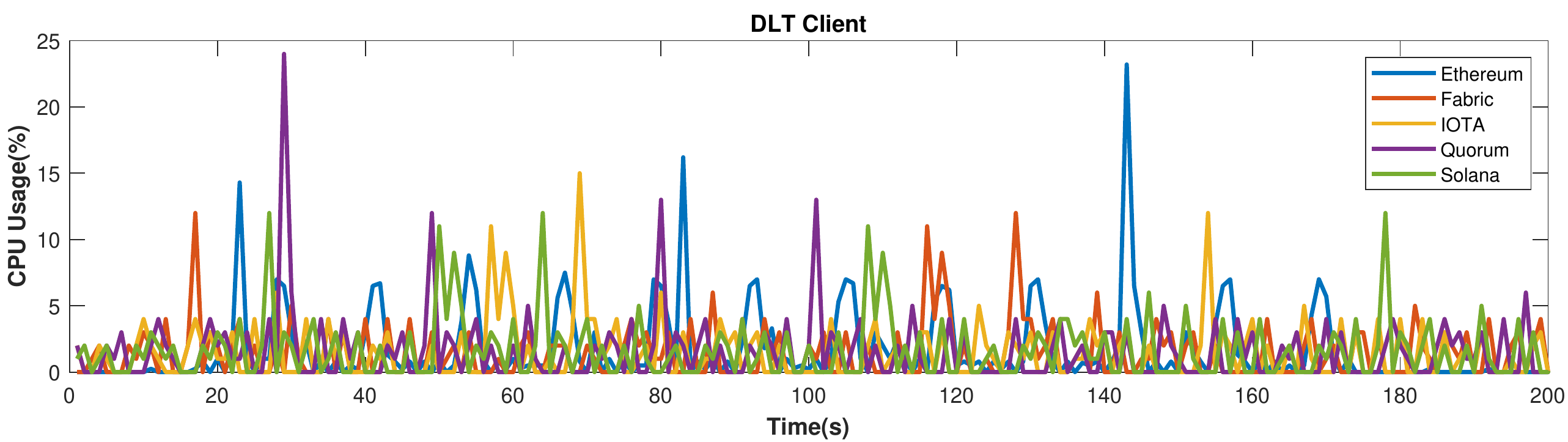}
         \caption{Energy consumption of \emph{DLT-clients} which transmit data to \emph{DLT-managers} and synchronize data when needed.}
         \label{fig:energy-client}
     \end{subfigure}
     \caption{Computation overhead of each network component of the 5 studied DLTs namely Ethereum, Hyperledger Fabric, IoTA, Quorum, and Solana.}
\end{figure}

During our evaluation, the DLT-client sends 10 transactions per second to the ledger, which is hosted by the DLT-manager-1 and -2. \lamrv{The reason why we set 10 transactions per second is that traffic from UR5 robots to a server is usually low at around 1-5 updates per hour or per day, depending on the specific scenario. Therefore, we stress the network with 10 TPS to test the efficiency of the system. In addition, we assume that the time required for completing a task with a robot could be a day or a week, so the robot could update the task progress and sensing information in a period of minutes or hours is sufficient.} By recording the CPU usage percentage of the DLT-specifc processes, we have observed the computation overhead in each case of the 5 selected DLT platforms. Looking at the DLT Managers, we have found Ethereum as an outlier, as it requires by far the most computation time of around \emph{85\%} of CPU as shown in Figure  \ref{fig:energy-managers} due to the usage of the Proof of Work (PoW) for the consensus and verification process. The \emph{non-PoW} DLTs, Solana, Hyperledger Fabric and Quorum, require in our private network setting only around \emph{1-3\%} CPU usage in both \emph{DLT Manager 1} and \emph{DLT Manager 2}. Similarly, the IOTA platform uses \emph{PoW} only rarely in order to prevent spam attacks, so the CPU usage of the DLT Managers is relatively low, similar to Hyperledger Fabric and Quorum. On the DLT Client component, the CPU usage is primarily the generation and transmission of transactions, so that these DLTs require around 5-10\% CPU usage of the Raspberry Pi as shown in Figure \ref{fig:energy-client}.

\begin{figure}[t!]
     \centering
     \begin{subfigure}[b]{0.45\textwidth}
         \centering
         \includegraphics[width=0.8\linewidth]{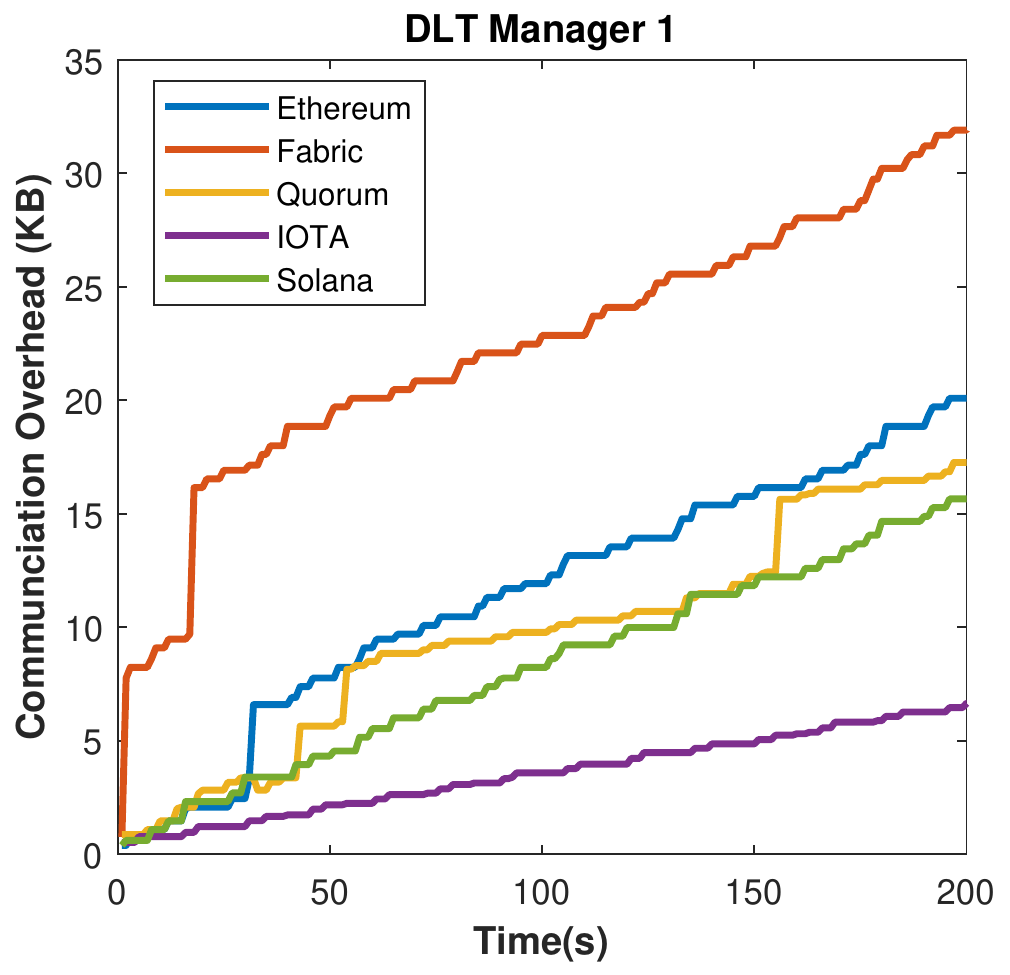}
         \caption{Communication overhead of \emph{DLT-manager-1}.}
         \label{fig:com-overhead1}
     \end{subfigure}
     \begin{subfigure}[b]{0.45\textwidth}
         \centering
         \includegraphics[width=0.8\linewidth]{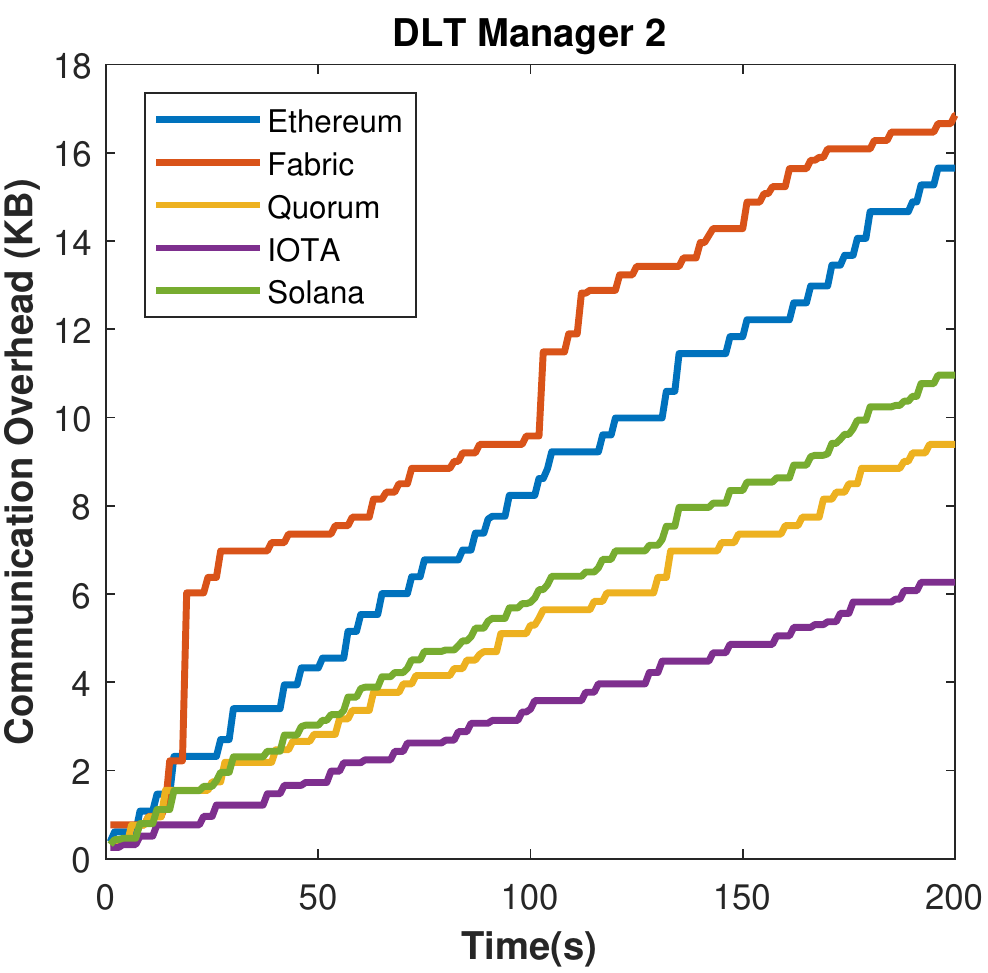}
         \caption{Communication overhead of \emph{DLT-manager-2}.}
         \label{fig:com-overhead2}
     \end{subfigure}
     \caption{Communication Overhead comparison of the 5 studied DLTs namely Ethereum, Hyperledger Fabric, IoTA, Quorum, and Solana.}
\end{figure}

\begin{figure}[t!]
    \centering
    \includegraphics[width=0.4\linewidth]{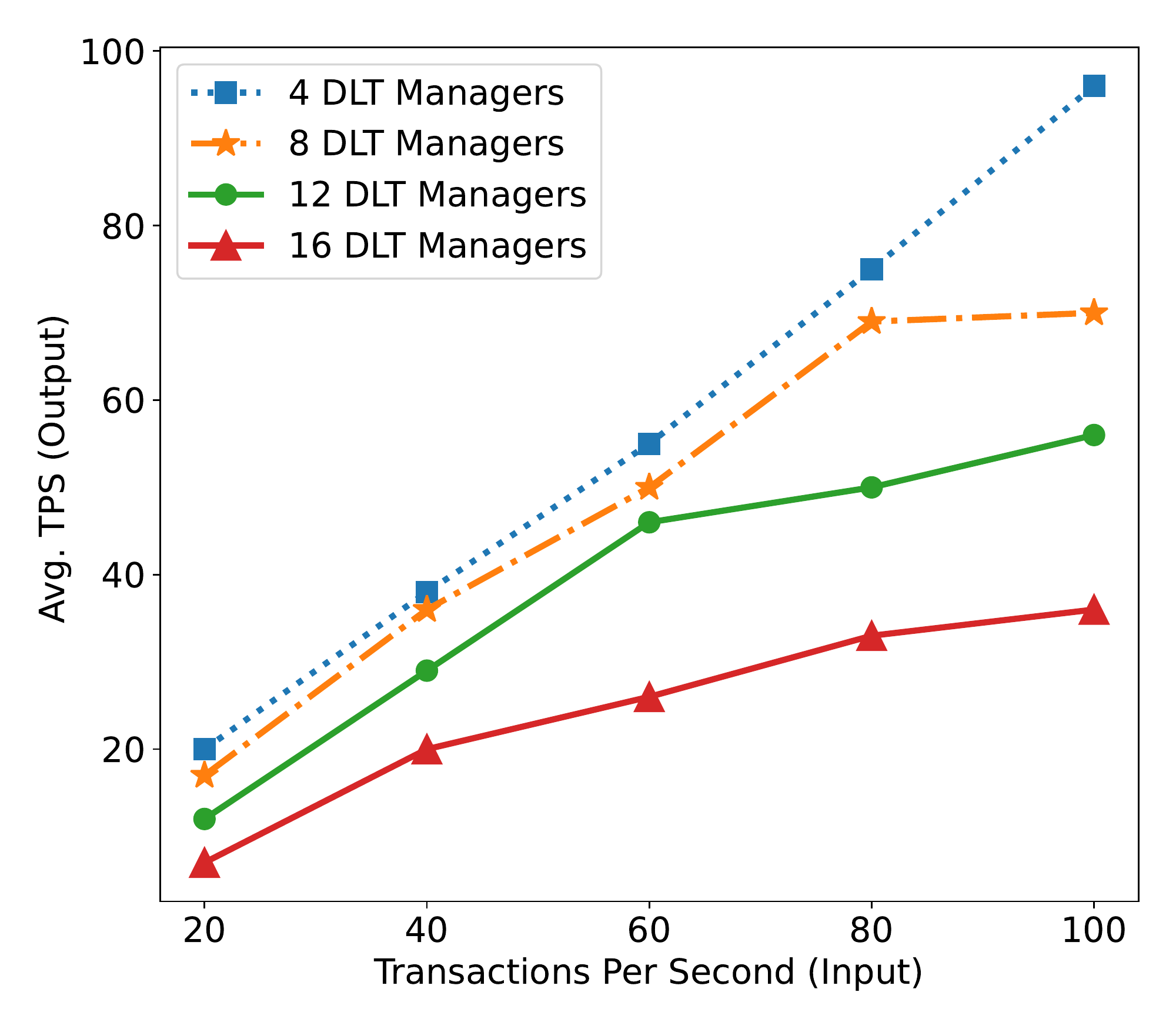}
    \caption{Impact of number of DLT-managers on the average transactions per second.}
    \label{fig:scalable-testing}
\end{figure}

Figure  \ref{fig:com-overhead1} and \ref{fig:com-overhead2} show the communication overhead of the five different DLTs in our shared manufacturing setup. The Hyperledger Fabric produces more network traffic than the others on the DLT Managers. The reason is that the network architecture of Hyperledger Fabric is optimized for an enterprise environment with high security requirements, where the raw data need to be formatted for signed transaction proposals, then going through the complex endorsement and validation process, before attachment to the Blockchain. This process introduces more communication overhead. IOTA produces the lowest traffic on the DLT Managers thanks to the design based on the Tangle\cite{popov2019iota}. Specifically, the interconnected Tangle infrastructure does not require total verification across the whole ledger. Instead, all parties are verifying simultaneously and, as a result, the energy and time required to complete transactions are shortened. In addition, Tangle's verification process purports to ensure that there are no duplicate transactions that would lead to double-spending.
On the \emph{DLT-clients}, the communication overhead is mainly coming from publishing the collected data via formatted transactions to DLT managers. In specific, a single transaction in IOTA consists of \emph{2673 trytes} which is equivalent to \emph{1589 bytes}, if encoded, a Ethereum transaction includes around \emph{109 bytes} of header, and no limited metadata %\todo{This is significantly less than IOTA. Is it correct? - yes}
, Hyperledger Fabric transaction sizes depends on the type of transactions, for example, \emph{3.06 kB} for spend and \emph{4.33 kB} for mint\cite{androulaki2018hyperledger}. 
%\todo{can you give a range?}. 
The overhead of a Solana transaction includes \emph{64 bytes} signature and  maximum \emph{1232} bytes for given metadata.

\lamrv{In order to analyze the scalability of the integration of Blockchains in manufacturing, we simulate Blockchain networks with a varying number of \textit{DLT managers} and \textit{DLT clients}, from $4$ to $16$ nodes. The number of input transactions per second is generated from 20 TPS to 100 TPS.  In this experiment, we choose to simulate the Ethereum Blockchain network. The simulation is extended from the framework \cite{strobel2020blockchain} of Blockchain for Swarm Robotics. Specifically, both \textit{ DLT-manager} and \textit{DLT-client} is executed in separated Docker Containers. The simulated robots are considered as light Blockchain clients and periodically published data to the managers for validation. The results in Fig. \ref{fig:scalable-testing} show that the more the number of \textit{DLT-managers} increases, the more the number of transactions validated per second decreases. The reason is that the more number of validators leading to the more number of transactions to be validated and exchanged among managers, as well as adding the system some delays in validating transactions. We observe that increasing the number of participants raises the challenges e.g, scalability, throughput, latency. Hence, private Blockchains are suitable for industrial IoT manufacturing environments. However, hybrid DLTs which provides flexibility on data visibility without compromising security also could be immense potential.}

Based on the CPU usage and the utilized computing hardware, we determined the energy consumption of the five selected DLT platforms and computed the carbon footprint. 
We assume that electricity for running the computational operations is consumed and produced in Germany. As a measure of carbon intensity of the German energy mix, calculated in a life cycle perspective, we used data from the life cycle database ecoinvent v.3 cutoff system model\cite{wernet2016ecoinvent}. In particular, the dataset \emph{Market for electricity, low voltage, DE} was chosen, which represents an average low-voltage energy mix for Germany. We obtained the life cycle impact of producing \emph{1 kWh} electricity according to this version of the database via the software \emph{SimaPro} and using the default IPCC Global Warming Potential (GWP) method with a time horizon of 100 years \cite{stocker2013ipcc}. This resulted in a value of global warming impact of \emph{0.540} \emph{kg \ch{CO2}-eq / kWh} that represents the impact of all greenhouse gases emitted in the  electricity production process and upstream activities in a life cycle perspective. This value was further used to calculate the total carbon footprint of the computation based on its energy requirements.
The results are shown in Table \ref{tab:co2}. We observe that the annual \emph{\ch{CO2}} generated through PoW consensus is significantly higher than that of non-PoW Blockchains. 
The private Ethereum DLT produced around \emph{26692 $\cdot 10^{-6}$} \emph{kg \ch{CO2}-eq/hour}, which is equivalent to the average of \emph{4.3} charged smartphones\cite{Greenhou56:online}. This compares to around \emph{203 $\cdot 10^{-6}$} \emph{kg\ch{CO2}-eq/hour}, \emph{211} \emph{kg\ch{CO2}-eq/hour}, and \emph{198} \emph{kg \ch{CO2}-eq/hour} from Hyperledger Fabric, Quorum, and IOTA, and Solana respectively. Note that all results are extrapolated to the utilization of our private DLT setup for the shared manufacturing application over an operation day.

%\input{table-co2}

%%%%%%%%%% CO2 Table %%%%

\begin{table} [h!]
\centering
\caption{Carbon FootPrint of private DLT testbed with 5 \emph{DLTs} calculuated in Germany market running per hour }
\begin{tabular}{p{1.7cm}||p{1.5cm}|p{2cm}|p{2cm}|p{2cm}|p{2cm}|p{2cm}}
\toprule
\textbf{Platform}  & \textbf{Power of Machine (kW)} & \textbf{Energy consumed on Average (kWh)} &  \textbf{Avg. CPU Usage for Blockchain Operation (\%)} & \textbf{Energy Consumed for Blockchain operation (kWh)} & \textbf{Greenhouse gas (GHG) emission in DE (kg \ch{CO2}-eq/kWh)}** & \textbf{GHG emission per blockchain operation (kg \ch{CO2}-eq)} \\
\midrule
\textit{Hyper.Fabric}        & 0.06  & 0.06 & 0.625\% & 375 $\cdot 10^{-6}$     & 0.540 & 203 $\cdot 10^{-6}$ \\
\textit{Ethereum }          & 0.06  & 0.06 & 82.35\% & 49392 $\cdot 10^{-6}$    & 0.540 & 26682 $\cdot 10^{-6}$  \\
\textit{Quorum}             & 0.06  & 0.06 & 0.65\% & 390 $\cdot 10^{-6}$       & 0.540 & 211 $\cdot 10^{-6}$ \\
\textit{IOTA}               & 0.06  & 0.06 & 0.61\% & 366 $\cdot 10^{-6}$       & 0.540 & 198 $\cdot 10^{-6}$ \\
\textit{Solana}             & 0.06  & 0.06 & 0.61\% & 366 $\cdot 10^{-6}$       & 0.540 & 198 $\cdot 10^{-6}$  \\
\bottomrule
\end{tabular}
%\begin{tablenotes}
%\centering
%      \small
%      \item *The Electronic Cost of \emph{DLT} nodes is computed using the PowerTop\cite{powertop71:online}
      %\item **Greenhouse gas (GHG) emission intensity of the Germany energy grid (kg CO2-eq /kWh)
%    \end{tablenotes}
\label{tab:co2}
\end{table}

%%%%%%%%%

\section*{Discussion}

We have seen that deploying a DLT in an industrial manufacturing environment allows to realize novel business cases. Choosing the right DLT platform for industrial use cases, such as the one we elaborated above, is challenging, as their are many options available. Therefore, we have conducted here an evaluation of five of the most popular and promising DLT platforms and proposed how to integrate those into a physical manufacturing system.

A clear observation is that Ethereum is an outlier in terms of CPU usage, due to its PoW consensus algorithm, which of course also results in high energy consumption. Therefore, we can conclude that Ethereum and other PoW DLTs should, in general, not be used in the envisioned manufacturing environments. \arne{In order to still be able to use many of Ethereum development tools,} a plant operator can use Quorum, which is an enterprise version of Ethereum. Both Quorum and Hyperledger Fabric show a similar performance in our local evaluation regarding CPU usage. However, Hyperledger Fabric introduces higher communication overhead as compared to Quorum. Therefore, in an environment with communication restrictions, the operator could opt for Quorum out of these two by simply looking at slight performance advantages. 
IOTA, which is specifically designed for IoT networks, requires the lowest CPU and communication overhead and can hence be favoured by an operator that has strong requirements in this regards. \arne{However, IOTA's smart contract mechanism is still under development and also the tooling support is not as strong}. Finally, Solana is a public Blockchain network with a focus on achieving high scalability. In our local network, Solana performed similarly to Quorum in terms of communication overhead as well as CPU usage. 

The results measured from our local experiments can be considered as a benchmark regarding sustainability aspects in specific shared manufacturing use cases. In the scope of this research, we evaluated the greenhouse gas emission per Blockchain operation based on energy consumed by Blockchain activities. For example, IOTA foundation provided an energy benchmark for the IOTA network, which shows results that are similar to our experimental results\cite{EnergyBe46:online}. In terms of Hyperledger Fabric, we have used Hyperledger Caliper for the benchmark  evaluation\cite{Hyperled13:online}. Referring to prior research\cite{platt2021energy}, the energy consumed by beyond-PoW blockchains, such as Polkadot\cite{wood2016polkadot}, Cardano\cite{CardanoH13:online}, or Hedera Hashgraph\cite{baird2018hedera}, is within a range that is similar to the energy consumed in our experimental setup. 

%Besides technical performance aspects, we also discuss the availability to adapt these DLTs on current manufacturing system. Ethereum - the world's second biggest crypto already is a major player in decentralized applications (dApps) after Bitcoin, has already had full distributed infrastructure and applications for customers making life a whole lot easier for users as well as developers. However, from the view of business administrators, they more often use a private DLTs to build multi-party business applications with high scalability within a trusted environment.

Looking towards the future, we see many benefits for the use of DLT in manufacturing, enabling a broad range of use cases and business models.
The vision at the horizon is a truly \emph{collaborative industrial IoT} in which \emph{things} (such as machines in a manufacturing plant) ubiquitously and automatically interact without intervention of humans. This is fuelled by the capability to autonomously make (micro-)payments. This would empower devices, e.g., to rent cloud server capacity for additional computational capacity when required, to pay directly to other devices for access to the Internet, or automatically pay for electricity consumed. The current payment systems are not well suited for massive-scale micro-transactions due to high transaction costs and limited capacity. This calls for a vision of payments between things, which will be a small per transaction, but autonomous and running efficiently at a massive scale. \lamrv{Besides, the research on governance, security aspects, and optimal consensus mechanism could be considered as a future work.}

%Also there is big challenge in using current credit-card systems that we have to share credit-card information with our device and allow it to share it with other devices for payments.

%\begin{enumerate}
%%\item concluding remarks on measurements 
%\item Level of support, code and API availability -> Ethereum has some advantages
%\item Support for enterprise level features (e.g., private channels)
%\item Activity of community (provide some measurements that support our statements, e.g., number of GIT commits per week, or number meetings, size of conference
%\item Future vision of thing to thing payments
%\end{enumerate}

%\todo{Add to discussion: is 10 TPS reasonable?}

\bibliography{main}

\begin{thebibliography}{10}
\urlstyle{rm}
\expandafter\ifx\csname url\endcsname\relax
  \def\url#1{\texttt{#1}}\fi
\expandafter\ifx\csname urlprefix\endcsname\relax\def\urlprefix{URL }\fi
\expandafter\ifx\csname doiprefix\endcsname\relax\def\doiprefix{DOI: }\fi
\providecommand{\bibinfo}[2]{#2}
\providecommand{\eprint}[2][]{\url{#2}}

\bibitem{kafle2021topcon}
\bibinfo{author}{Kafle, B.} \emph{et~al.}
\newblock \bibinfo{journal}{\bibinfo{title}{Topcon--technology options for cost
  efficient industrial manufacturing}}.
\newblock {\emph{\JournalTitle{Solar Energy Materials and Solar Cells}}}
  \textbf{\bibinfo{volume}{227}}, \bibinfo{pages}{111100}
  (\bibinfo{year}{2021}).

\bibitem{soret2021learning}
\bibinfo{author}{Soret, B.} \emph{et~al.}
\newblock \bibinfo{journal}{\bibinfo{title}{Learning, computing, and
  trustworthiness in intelligent iot environments: Performance-energy
  tradeoffs}}.
\newblock {\emph{\JournalTitle{IEEE Transactions on Green Communications and
  Networking}}}  (\bibinfo{year}{2021}).

\bibitem{helu2015identifying}
\bibinfo{author}{Helu, M.}, \bibinfo{author}{Morris, K.},
  \bibinfo{author}{Jung, K.}, \bibinfo{author}{Lyons, K.} \&
  \bibinfo{author}{Leong, S.}
\newblock \bibinfo{journal}{\bibinfo{title}{Identifying performance assurance
  challenges for smart manufacturing}}.
\newblock {\emph{\JournalTitle{Manufacturing letters}}}
  \textbf{\bibinfo{volume}{6}}, \bibinfo{pages}{1--4} (\bibinfo{year}{2015}).

\bibitem{al2019blockchain}
\bibinfo{author}{Al-Jaroodi, J.} \& \bibinfo{author}{Mohamed, N.}
\newblock \bibinfo{journal}{\bibinfo{title}{Blockchain in industries: A
  survey}}.
\newblock {\emph{\JournalTitle{IEEE Access}}} \textbf{\bibinfo{volume}{7}},
  \bibinfo{pages}{36500--36515} (\bibinfo{year}{2019}).

\bibitem{tschorsch2016bitcoin}
\bibinfo{author}{Tschorsch, F.} \& \bibinfo{author}{Scheuermann, B.}
\newblock \bibinfo{journal}{\bibinfo{title}{Bitcoin and beyond: A technical
  survey on decentralized digital currencies}}.
\newblock {\emph{\JournalTitle{IEEE Communications Surveys \& Tutorials}}}
  \textbf{\bibinfo{volume}{18}}, \bibinfo{pages}{2084--2123}
  (\bibinfo{year}{2016}).

\bibitem{alrebdi2022svbe}
\bibinfo{author}{Alrebdi, N.}, \bibinfo{author}{Alabdulatif, A.},
  \bibinfo{author}{Iwendi, C.} \& \bibinfo{author}{Lian, Z.}
\newblock \bibinfo{journal}{\bibinfo{title}{Svbe: searchable and verifiable
  blockchain-based electronic medical records system}}.
\newblock {\emph{\JournalTitle{Scientific Reports}}}
  \textbf{\bibinfo{volume}{12}}, \bibinfo{pages}{1--11} (\bibinfo{year}{2022}).

\bibitem{Top10Sec35:online}
\bibinfo{title}{Top 10 security predictions 2016}.
\newblock
  \bibinfo{howpublished}{\url{https://www.gartner.com/smarterwithgartner/top-10-security-predictions-2016}}.
\newblock \bibinfo{note}{(Accessed on 12/06/2021)}.

\bibitem{nguyen2021marketplace}
\bibinfo{author}{Nguyen, L.~D.}, \bibinfo{author}{Pandey, S.~R.},
  \bibinfo{author}{Beatriz, S.}, \bibinfo{author}{Broering, A.} \&
  \bibinfo{author}{Popovski, P.}
\newblock \bibinfo{journal}{\bibinfo{title}{A marketplace for trading ai models
  based on blockchain and incentives for iot data}}.
\newblock {\emph{\JournalTitle{arXiv preprint arXiv:2112.02870}}}
  (\bibinfo{year}{2021}).

\bibitem{chen2018edge}
\bibinfo{author}{Chen, B.} \emph{et~al.}
\newblock \bibinfo{journal}{\bibinfo{title}{Edge computing in iot-based
  manufacturing}}.
\newblock {\emph{\JournalTitle{IEEE Communications Magazine}}}
  \textbf{\bibinfo{volume}{56}}, \bibinfo{pages}{103--109}
  (\bibinfo{year}{2018}).

\bibitem{hassan2019current}
\bibinfo{author}{Hassan, W.~H.} \emph{et~al.}
\newblock \bibinfo{journal}{\bibinfo{title}{Current research on internet of
  things (iot) security: A survey}}.
\newblock {\emph{\JournalTitle{Computer networks}}}
  \textbf{\bibinfo{volume}{148}}, \bibinfo{pages}{283--294}
  (\bibinfo{year}{2019}).

\bibitem{shi2016edge}
\bibinfo{author}{Shi, W.}, \bibinfo{author}{Cao, J.}, \bibinfo{author}{Zhang,
  Q.}, \bibinfo{author}{Li, Y.} \& \bibinfo{author}{Xu, L.}
\newblock \bibinfo{journal}{\bibinfo{title}{Edge computing: Vision and
  challenges}}.
\newblock {\emph{\JournalTitle{IEEE internet of things journal}}}
  \textbf{\bibinfo{volume}{3}}, \bibinfo{pages}{637--646}
  (\bibinfo{year}{2016}).

\bibitem{Economyo60:online}
\bibinfo{title}{Economy of things | bosch global}.
\newblock
  \bibinfo{howpublished}{\url{https://www.bosch.com/research/know-how/success-stories/economy-of-things-a-technology-and-business-evolution/}}.
\newblock \bibinfo{note}{(Accessed on 12/06/2021)}.

\bibitem{Blockcha24:online}
\bibinfo{title}{Blockchain iot | exclusive content for the food and beverage
  industry | siemens global}.
\newblock
  \bibinfo{howpublished}{\url{https://new.siemens.com/global/en/markets/food-beverage/exclusive-area/blockchain-iot.html}}.
\newblock \bibinfo{note}{(Accessed on 12/06/2021)}.

\bibitem{li2018toward}
\bibinfo{author}{Li, Z.}, \bibinfo{author}{Barenji, A.~V.} \&
  \bibinfo{author}{Huang, G.~Q.}
\newblock \bibinfo{journal}{\bibinfo{title}{Toward a blockchain cloud
  manufacturing system as a peer to peer distributed network platform}}.
\newblock {\emph{\JournalTitle{Robotics and computer-integrated
  manufacturing}}} \textbf{\bibinfo{volume}{54}}, \bibinfo{pages}{133--144}
  (\bibinfo{year}{2018}).

\bibitem{8671694}
\bibinfo{author}{Danzi, P.}, \bibinfo{author}{Kalør, A.~E.},
  \bibinfo{author}{Stefanović, C.} \& \bibinfo{author}{Popovski, P.}
\newblock \bibinfo{journal}{\bibinfo{title}{Delay and communication tradeoffs
  for blockchain systems with lightweight iot clients}}.
\newblock {\emph{\JournalTitle{IEEE Internet of Things Journal}}}
  \textbf{\bibinfo{volume}{6}}, \bibinfo{pages}{2354--2365},
  \doiprefix\url{10.1109/JIOT.2019.2906615} (\bibinfo{year}{2019}).

\bibitem{9129732}
\bibinfo{author}{Fan, C.}, \bibinfo{author}{Ghaemi, S.},
  \bibinfo{author}{Khazaei, H.} \& \bibinfo{author}{Musilek, P.}
\newblock \bibinfo{journal}{\bibinfo{title}{Performance evaluation of
  blockchain systems: A systematic survey}}.
\newblock {\emph{\JournalTitle{IEEE Access}}} \textbf{\bibinfo{volume}{8}},
  \bibinfo{pages}{126927--126950}, \doiprefix\url{10.1109/ACCESS.2020.3006078}
  (\bibinfo{year}{2020}).

\bibitem{fu2018blockchain}
\bibinfo{author}{Fu, B.}, \bibinfo{author}{Shu, Z.} \& \bibinfo{author}{Liu,
  X.}
\newblock \bibinfo{journal}{\bibinfo{title}{Blockchain enhanced emission
  trading framework in fashion apparel manufacturing industry}}.
\newblock {\emph{\JournalTitle{Sustainability}}} \textbf{\bibinfo{volume}{10}},
  \bibinfo{pages}{1105} (\bibinfo{year}{2018}).

\bibitem{yu2020blockchain}
\bibinfo{author}{Yu, C.}, \bibinfo{author}{Zhang, L.}, \bibinfo{author}{Zhao,
  W.} \& \bibinfo{author}{Zhang, S.}
\newblock \bibinfo{journal}{\bibinfo{title}{A blockchain-based service
  composition architecture in cloud manufacturing}}.
\newblock {\emph{\JournalTitle{International Journal of Computer Integrated
  Manufacturing}}} \textbf{\bibinfo{volume}{33}}, \bibinfo{pages}{701--715}
  (\bibinfo{year}{2020}).

\bibitem{androulaki2018hyperledger}
\bibinfo{author}{Androulaki, E.} \emph{et~al.}
\newblock \bibinfo{title}{Hyperledger fabric: a distributed operating system
  for permissioned blockchains}.
\newblock In \emph{\bibinfo{booktitle}{Proceedings of the thirteenth EuroSys
  conference}}, \bibinfo{pages}{1--15} (\bibinfo{year}{2018}).

\bibitem{baliga2018performance}
\bibinfo{author}{Baliga, A.}, \bibinfo{author}{Subhod, I.},
  \bibinfo{author}{Kamat, P.} \& \bibinfo{author}{Chatterjee, S.}
\newblock \bibinfo{journal}{\bibinfo{title}{Performance evaluation of the
  quorum blockchain platform}}.
\newblock {\emph{\JournalTitle{arXiv preprint arXiv:1809.03421}}}
  (\bibinfo{year}{2018}).

\bibitem{vujivcic2018blockchain}
\bibinfo{author}{Vuji{\v{c}}i{\'c}, D.}, \bibinfo{author}{Jagodi{\'c}, D.} \&
  \bibinfo{author}{Ran{\dj}i{\'c}, S.}
\newblock \bibinfo{title}{Blockchain technology, bitcoin, and ethereum: A brief
  overview}.
\newblock In \emph{\bibinfo{booktitle}{2018 17th international symposium
  infoteh-jahorina (infoteh)}}, \bibinfo{pages}{1--6}
  (\bibinfo{organization}{IEEE}, \bibinfo{year}{2018}).

\bibitem{popov2019iota}
\bibinfo{author}{Popov, S.} \& \bibinfo{author}{Lu, Q.}
\newblock \bibinfo{journal}{\bibinfo{title}{Iota: feeless and free}}.
\newblock {\emph{\JournalTitle{IEEE Blockchain Technical Briefs}}}
  (\bibinfo{year}{2019}).

\bibitem{yakovenko2018solana}
\bibinfo{author}{Yakovenko, A.}
\newblock \bibinfo{journal}{\bibinfo{title}{Solana: A new architecture for a
  high performance blockchain v0. 8.13}}.
\newblock {\emph{\JournalTitle{Whitepaper}}}  (\bibinfo{year}{2018}).

\bibitem{sguanci2021layer}
\bibinfo{author}{Sguanci, C.}, \bibinfo{author}{Spatafora, R.} \&
  \bibinfo{author}{Vergani, A.~M.}
\newblock \bibinfo{journal}{\bibinfo{title}{Layer 2 blockchain scaling: a
  survey}}.
\newblock {\emph{\JournalTitle{arXiv preprint arXiv:2107.10881}}}
  (\bibinfo{year}{2021}).

\bibitem{wust2018you}
\bibinfo{author}{W{\"u}st, K.} \& \bibinfo{author}{Gervais, A.}
\newblock \bibinfo{title}{Do you need a blockchain?}
\newblock In \emph{\bibinfo{booktitle}{2018 Crypto Valley Conference on
  Blockchain Technology (CVCBT)}}, \bibinfo{pages}{45--54}
  (\bibinfo{organization}{IEEE}, \bibinfo{year}{2018}).

\bibitem{boneh2018verifiable}
\bibinfo{author}{Boneh, D.}, \bibinfo{author}{Bonneau, J.},
  \bibinfo{author}{B{\"u}nz, B.} \& \bibinfo{author}{Fisch, B.}
\newblock \bibinfo{title}{Verifiable delay functions}.
\newblock In \emph{\bibinfo{booktitle}{Annual international cryptology
  conference}}, \bibinfo{pages}{757--788} (\bibinfo{organization}{Springer},
  \bibinfo{year}{2018}).

\bibitem{IOTASmar41:online}
\bibinfo{title}{Iota smart contracts beta release}.
\newblock
  \bibinfo{howpublished}{\url{https://blog.iota.org/iota-smart-contracts-beta-release/}}.
\newblock \bibinfo{note}{(Accessed on 05/04/2022)}.

\bibitem{Homepage72:online}
\bibinfo{title}{Homepage | solana docs}.
\newblock \bibinfo{howpublished}{\url{https://docs.solana.com/}}.
\newblock \bibinfo{note}{(Accessed on 01/14/2022)}.

\bibitem{xiao2020survey}
\bibinfo{author}{Xiao, Y.}, \bibinfo{author}{Zhang, N.}, \bibinfo{author}{Lou,
  W.} \& \bibinfo{author}{Hou, Y.~T.}
\newblock \bibinfo{journal}{\bibinfo{title}{A survey of distributed consensus
  protocols for blockchain networks}}.
\newblock {\emph{\JournalTitle{IEEE Communications Surveys \& Tutorials}}}
  \textbf{\bibinfo{volume}{22}}, \bibinfo{pages}{1432--1465}
  (\bibinfo{year}{2020}).

\bibitem{ahl2020exploring}
\bibinfo{author}{Ahl, A.} \emph{et~al.}
\newblock \bibinfo{journal}{\bibinfo{title}{Exploring blockchain for the energy
  transition: Opportunities and challenges based on a case study in japan}}.
\newblock {\emph{\JournalTitle{Renewable and sustainable energy reviews}}}
  \textbf{\bibinfo{volume}{117}}, \bibinfo{pages}{109488}
  (\bibinfo{year}{2020}).

\bibitem{xiong2018mobile}
\bibinfo{author}{Xiong, Z.}, \bibinfo{author}{Zhang, Y.},
  \bibinfo{author}{Niyato, D.}, \bibinfo{author}{Wang, P.} \&
  \bibinfo{author}{Han, Z.}
\newblock \bibinfo{journal}{\bibinfo{title}{When mobile blockchain meets edge
  computing}}.
\newblock {\emph{\JournalTitle{IEEE Communications Magazine}}}
  \textbf{\bibinfo{volume}{56}}, \bibinfo{pages}{33--39}
  (\bibinfo{year}{2018}).

\bibitem{UR5colla78:online}
\bibinfo{title}{Ur5 collaborative robot arm | flexible and lightweight cobot}.
\newblock
  \bibinfo{howpublished}{\url{https://www.universal-robots.com/products/ur5-robot/}}.
\newblock \bibinfo{note}{(Accessed on 01/26/2022)}.

\bibitem{nguyen2021modeling}
\bibinfo{author}{Nguyen, L.~D.}, \bibinfo{author}{Leyva-Mayorga, I.},
  \bibinfo{author}{Lewis, A.~N.} \& \bibinfo{author}{Popovski, P.}
\newblock \bibinfo{journal}{\bibinfo{title}{Modeling and analysis of data
  trading on blockchain-based market in iot networks}}.
\newblock {\emph{\JournalTitle{IEEE Internet of Things Journal}}}
  \textbf{\bibinfo{volume}{8}}, \bibinfo{pages}{6487--6497}
  (\bibinfo{year}{2021}).

\bibitem{storch2021incentive}
\bibinfo{author}{Storch, D.-M.}, \bibinfo{author}{Timme, M.} \&
  \bibinfo{author}{Schr{\"o}der, M.}
\newblock \bibinfo{journal}{\bibinfo{title}{Incentive-driven transition to high
  ride-sharing adoption}}.
\newblock {\emph{\JournalTitle{Nature communications}}}
  \textbf{\bibinfo{volume}{12}}, \bibinfo{pages}{1--10} (\bibinfo{year}{2021}).

\bibitem{RentingM65:online}
\bibinfo{title}{Renting machine made easy – mcpond}.
\newblock \bibinfo{howpublished}{\url{https://mcpond.com/}}.
\newblock \bibinfo{note}{(Accessed on 12/07/2021)}.

\bibitem{jourenko2019sok}
\bibinfo{author}{Jourenko, M.}, \bibinfo{author}{Kurazumi, K.},
  \bibinfo{author}{Larangeira, M.} \& \bibinfo{author}{Tanaka, K.}
\newblock \bibinfo{journal}{\bibinfo{title}{Sok: A taxonomy for layer-2
  scalability related protocols for cryptocurrencies.}}
\newblock {\emph{\JournalTitle{IACR Cryptol. ePrint Arch.}}}
  \textbf{\bibinfo{volume}{2019}}, \bibinfo{pages}{352} (\bibinfo{year}{2019}).

\bibitem{strobel2020blockchain}
\bibinfo{author}{Strobel, V.}, \bibinfo{author}{Castell{\'o}~Ferrer, E.} \&
  \bibinfo{author}{Dorigo, M.}
\newblock \bibinfo{journal}{\bibinfo{title}{Blockchain technology secures robot
  swarms: A comparison of consensus protocols and their resilience to byzantine
  robots}}.
\newblock {\emph{\JournalTitle{Frontiers in Robotics and AI}}}
  \textbf{\bibinfo{volume}{7}}, \bibinfo{pages}{54} (\bibinfo{year}{2020}).

\bibitem{wernet2016ecoinvent}
\bibinfo{author}{Wernet, G.} \emph{et~al.}
\newblock \bibinfo{journal}{\bibinfo{title}{The ecoinvent database version 3
  (part i): overview and methodology}}.
\newblock {\emph{\JournalTitle{The International Journal of Life Cycle
  Assessment}}} \textbf{\bibinfo{volume}{21}}, \bibinfo{pages}{1218--1230}
  (\bibinfo{year}{2016}).

\bibitem{stocker2013ipcc}
\bibinfo{author}{Stocker, T.~F.} \emph{et~al.}
\newblock \bibinfo{title}{Ipcc, 2013: climate change 2013: the physical science
  basis. contribution of working group i to the fifth assessment report of the
  intergovernmental panel on climate change} (\bibinfo{year}{2013}).

\bibitem{Greenhou56:online}
\bibinfo{title}{Greenhouse gas equivalencies calculator | us epa}.
\newblock
  \bibinfo{howpublished}{\url{https://www.epa.gov/energy/greenhouse-gas-equivalencies-calculator}}.
\newblock \bibinfo{note}{(Accessed on 12/07/2021)}.

\bibitem{EnergyBe46:online}
\bibinfo{title}{Energy benchmarks for the iota network (chrysalis edition)}.
\newblock
  \bibinfo{howpublished}{\url{https://blog.iota.org/internal-energy-benchmarks-for-iota/}}.
\newblock \bibinfo{note}{(Accessed on 02/11/2022)}.

\bibitem{Hyperled13:online}
\bibinfo{title}{Hyperledger caliper – hyperledger foundation}.
\newblock
  \bibinfo{howpublished}{\url{https://www.hyperledger.org/use/caliper}}.
\newblock \bibinfo{note}{(Accessed on 02/11/2022)}.

\bibitem{platt2021energy}
\bibinfo{author}{Platt, M.} \emph{et~al.}
\newblock \bibinfo{journal}{\bibinfo{title}{Energy footprint of blockchain
  consensus mechanisms beyond proof-of-work}}.
\newblock {\emph{\JournalTitle{arXiv preprint arXiv:2109.03667}}}
  (\bibinfo{year}{2021}).

\bibitem{wood2016polkadot}
\bibinfo{author}{Wood, G.}
\newblock \bibinfo{journal}{\bibinfo{title}{Polkadot: Vision for a
  heterogeneous multi-chain framework}}.
\newblock {\emph{\JournalTitle{White Paper}}} \textbf{\bibinfo{volume}{21}},
  \bibinfo{pages}{2327--4662} (\bibinfo{year}{2016}).

\bibitem{CardanoH13:online}
\bibinfo{title}{Cardano | home}.
\newblock \bibinfo{howpublished}{\url{https://cardano.org/}}.
\newblock \bibinfo{note}{(Accessed on 02/11/2022)}.

\bibitem{baird2018hedera}
\bibinfo{author}{Baird, L.}, \bibinfo{author}{Harmon, M.} \&
  \bibinfo{author}{Madsen, P.}
\newblock \bibinfo{journal}{\bibinfo{title}{Hedera: A governing council \&
  public hashgraph network}}.
\newblock {\emph{\JournalTitle{The trust layer of the internet, whitepaper}}}
  \textbf{\bibinfo{volume}{1}}, \bibinfo{pages}{1--97} (\bibinfo{year}{2018}).

\end{thebibliography}

\section*{Acknowledgements}

This work has received funding from the European Union’s Horizon 2020 research and innovation programme under grant agreement No. 957218 (Project IntellIoT).

\section*{Author contributions statement}
L.D.N and A.B proposed the idea, implemented the system, and drafted the main manuscript. P.P revised manuscript and derived the structure for the research. M.P analyzed the sustainability aspect regarding \ch{CO2} emission of the research.
All authors contributed, edited, and reviewed the manuscript. 

\section*{Competing interests}
The authors declare no competing interests

%Must include all authors, identified by initials, for example:
%A.A. conceived the experiment(s),  A.A. and B.A. conducted the experiment(s), C.A. and D.A. analysed the results.  All authors reviewed the manuscript. 

%\section*{Additional information}

%To include, in this order: \textbf{Accession codes} (where applicable); \textbf{Competing interests} (mandatory statement). 

%The corresponding author is responsible for submitting a \href{http://www.nature.com/srep/policies/index.html#competing}{competing interests statement} on behalf of all authors of the paper. This statement must be included in the submitted article file.

%https://new.siemens.com/global/en/products/financing/whitepapers/whitepaper-financing-decarbonization/manufacturing.html

\section*{Data Availability}
The datasets used and/or analysed during the current study available from the corresponding author on reasonable request.

\end{document}